\pdfoutput=1
\PassOptionsToPackage{bookmarks=false,colorlinks=true,linkcolor=blue,urlcolor=cyan,filecolor=magenta,citecolor=red,pdfstartview=FitV,hyperfootnotes=false,pdftitle={Carroll in Shallow Water},pdfauthor={Arjun Bagchi, Aritra Banerjee, Saikat Mondal, Sayantan Sarkar},pdfsubject={Carrollinwater},pdfkeywords={carrollian symmetries and shallow water waves},bookmarksopen=true}{hyperref}
\documentclass[aps,prl,reprint,superscriptaddress]{revtex4-2} 

\usepackage[dvipsnames]{xcolor}
\usepackage[bookmarks=false,colorlinks=true,linkcolor=blue,urlcolor=cyan,filecolor=magenta,citecolor=red,pdfstartview=FitV,hyperfootnotes=false,pdftitle={Carroll in Shallow Water},pdfauthor={Arjun Bagchi, Aritra Banerjee, Saikat Mondal, Sayantan Sarkar},pdfsubject={Carrollinwater},pdfkeywords={carrollian symmetries and shallow water waves},bookmarksopen=true]{hyperref}
\usepackage{amssymb,amsmath,amsfonts}
\usepackage{epsfig}
\usepackage{graphicx}
\usepackage{epstopdf}
\usepackage{natbib}
\usepackage{braket}
\usepackage{dutchcal}
\usepackage{comment}

\newcommand{\C}{\mathbb{C}}

\newcommand{\be}{\begin{eqnarray}}
	\newcommand{\ee}{\end{eqnarray}}
\newcommand{\beq}{\begin{eqnarray}}
	\newcommand{\eeq}{\end{eqnarray}}

\newcommand{\beqa}{\begin{eqnarray}}
	\newcommand{\eeqa}{\end{eqnarray}}

\usepackage{mathrsfs}

\renewcommand{\sl}{\mathfrak{sl}}

\newcommand{\x}{\boldsymbol{x}}
\newcommand{\y}{\boldsymbol{y}}

\ifdefined\C
\renewcommand{\C}{\boldsymbol{C}}
\else
\newcommand{\C}{\boldsymbol{C}}
\fi
\renewcommand{\L}{\mathcal{L}}

\definecolor{gris}{rgb}{0.5,0.5,0.5}
\definecolor{darkgreen}{rgb}{0.0,0.5,0.0}

\allowdisplaybreaks[0]
\begin{document}

		
	\title{Carroll in Shallow Water}
	
	\author{Arjun Bagchi}
	\email{abagchi@iitk.ac.in}
	\affiliation{Indian Institute of Technology Kanpur, Kanpur 208016, India}
	
\author{Aritra Banerjee} 
\email{aritra.banerjee@pilani.bits-pilani.ac.in}
	\affiliation{Birla Institute of Technology and Science, Pilani Campus, Rajasthan 333031, India}
 \author{Saikat Mondal} 
 \email{saikatmd@iitk.ac.in}
	\affiliation{Indian Institute of Technology Kanpur, Kanpur 208016, India}
 \author{Sayantan Sarkar} 
 \email{sayantans22@iitk.ac.in}
	\affiliation{Indian Institute of Technology Kanpur, Kanpur 208016, India}

\begin{abstract}
We discover a surprising connection between Carrollian symmetries and hydrodynamics in the shallow water approximation. Carrollian symmetries arise in the speed of light going to zero limit of relativistic Poincaré symmetries. Using a recent gauge theoretic description of shallow water wave equations we find that the actions corresponding to two different waves, viz. the so called flat band solution and the Poincaré waves map exactly to the actions of the electric and magnetic sectors of Carrollian electrodynamics. 
\end{abstract}
		
	\maketitle

 
\paragraph{Introduction.} 
Our daily experiences are deeply rooted in non-relativistic physics. The world around us, to a good approximation, is  mostly Galilean. Fundamental physics, however, is built on relativistically covariant theories. Symmetries pave the way to understanding of fundamental and effective theories, with the language to connect to symmetries being group theory. In terms of group theory, Galilean symmetries emerge from relativistic symmetries when one considers an In{\"o}n{\"u}-Wigner contraction of the Poincare group by sending the speed of light $c$ to infinity. This singular limit abelianizes the Lorentz boosts and make them Galilean. 

There exists another contraction of the Poincare algebra, where instead of sending the speed of light to infinity, one sends it to zero. The resulting group is called the Carroll group \cite{SenGupta:1966qer, LBLL} and it has many weird properties. Time was absolute in the Galilean world and space was relative. This flips around in the crazy world of Carrollian physics, -- space becomes absolute and time relative. 

One would wonder what bizarre systems would have Carrollian symmetry associate with them and for long these symmetries were considered a mere mathematical curiosity and cast aside. Of late however, there has been an avalanche of activities relating Carrollian symmetries to various physical systems of interest. In particular, Carrollian symmetry have been shown to arise in condensed matter systems like fractons \cite{Bidussi:2021nmp}, and systems with flat bands \cite{Bagchi:2022eui}, where some other effective velocity scale (like the Fermi velocity) goes to zero. The Carroll algebra also plays a starring role in hydrodynamics of the quark-gluon-plasma \cite{Bagchi:2023ysc,Bagchi:2023rwd}, in cosmology \cite{deBoer:2021jej} and on the event horizons of generic black holes \cite{Henneaux:1979vn,Donnay:2019jiz} as well as on all null hypersurfaces \cite{Ciambelli:2019lap}. Conformal extensions of Carrollian symmetries have been proposed as putative holographic duals of asymptotically flat spacetimes \cite{Bagchi:2010zz,Bagchi:2012cy,Bagchi:2012xr,Barnich:2012xq,Bagchi:2016bcd,Donnay:2022aba,Bagchi:2022emh,Donnay:2022wvx, Bagchi:2023fbj} and these symmetries also appear on the worldsheet of tensionless null strings \cite{Bagchi:2013bga,Bagchi:2019cay,Bagchi:2020fpr}. 

In this paper, we discover another, even more surprising, appearance of Carroll structures, this time in \textit{shallow water waves}. Shallow water dynamics, by definition, is applicable to fluid systems, specifically to a fluid layer of constant density, for which height is much smaller compared to its horizontal span. The dynamics of such fluid systems can be entirely determined by the momentum and mass continuity equations. Oceans and the atmosphere are examples of such systems. We will show that linearised theories of shallow water waves can be directly related to a Carrollian theory of electrodynamics in $d=3$, with consistent mapping of field variables.

In a recent and rather beautiful paper \cite{Tong:2022gpg}, David Tong showed that the non-linear shallow water equations can be recast as a (2+1) dimensional Abelian gauge theory, where the electric and magnetic field corresponds to conserved height and conserved vorticity of the fluid. In the linearised approximation, the solutions can be separated into the geostrophic flat band and Poincaré waves and for both of these solutions, effective actions have been constructed. We will show that these effective actions are exactly the ones of the so called \textit{Electric} and \textit{Magnetic} versions of Carrollian electrodynamics \cite{Duval:2014uoa}. 

\begin{center}
\begin{figure}
\includegraphics[scale=0.17]{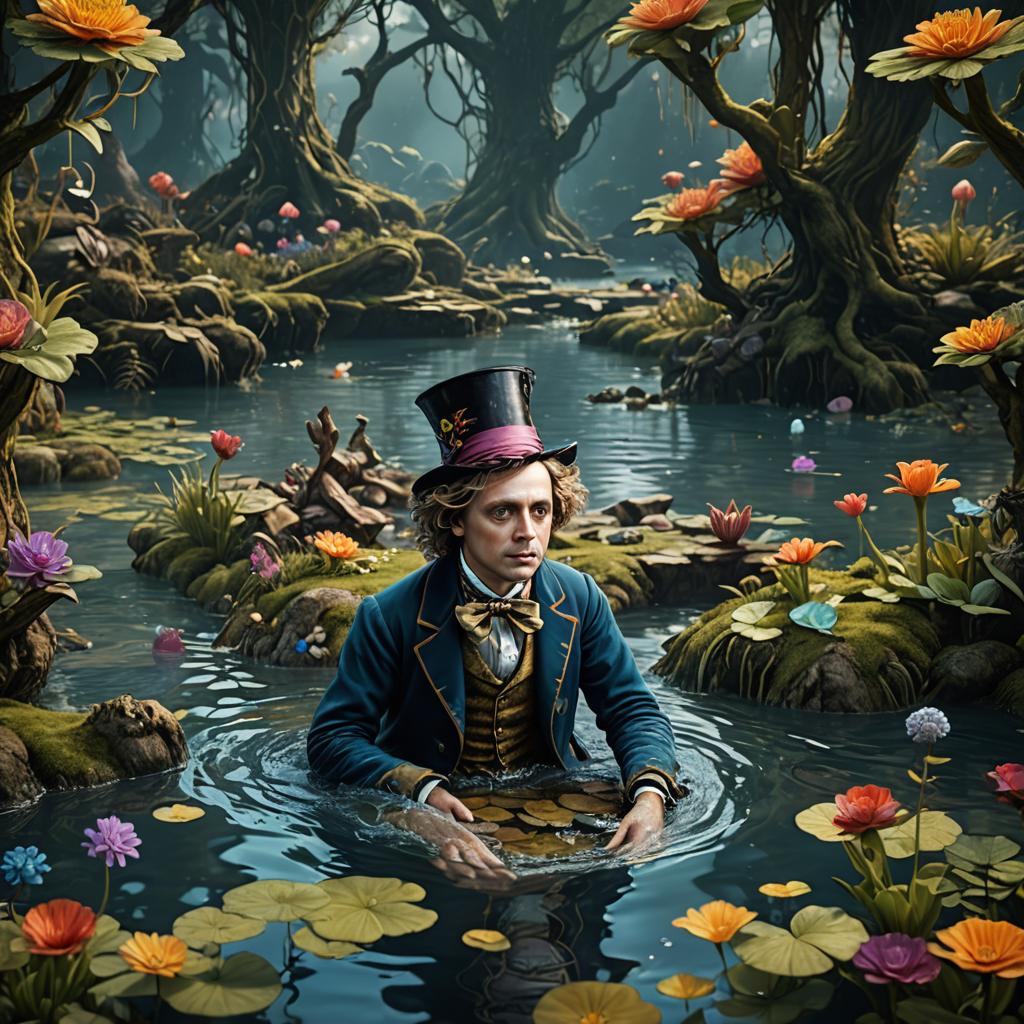}
\caption{Lewis Carroll in shallow water: AI generated image}
\end{figure}
\end{center}

\paragraph{Shallow water waves.} Consider fluids in $d=2+1$ dimensions. Here, following \cite{Tong:2022gpg}, one can describe the non-linear shallow water system completely by two dynamical fields: $h(x^i,t)$, which measures the height of the fluid and $\vec{u}(x^i,t)$ describing the horizontal fluid velocity, with $i=1,2$. In the Eulerian description, the governing equations of motion (EOM) are
    \begin{eqnarray}
        \frac{Dh}{Dt}=-h\nabla.\Vec{u}, \quad \frac{Du^i}{Dt}=f\epsilon^{ij}u^j-g\frac{\partial h}{\partial x^i}, \,  i=1,2. \label{eq:1}
    \end{eqnarray}
Here, the material derivative, $\frac{D}{Dt}=\frac{\partial }{\partial t}+\Vec{u}.\nabla$ is used to monitor the change in height in relation to the flow,
with $g$ being the gravitational constant and $f$ is the Coriolis parameter. The theory admits two conserved charges, which are given by zero modes of the currents,
\begin{eqnarray}\label{concharge}
    J^0=h, \, \Vec{J}=h\Vec{u}; \,\quad  \Tilde{J}^0=\zeta+f, \, {\Vec{\Tilde{J}}}=(\zeta+f)\Vec{u}.
\end{eqnarray} 
Here $\zeta=\epsilon^{ij}\partial_i u_j$ is the vorticity of the shallow water. 

\medskip

\paragraph{Nonlinear gauge theory.}
Following Tong \cite{Tong:2022gpg}, we now discuss the $u(1)\times u(1)$ gauge theoretic interpretation of shallow water theory. 
 Denoting 2+1 dimensional gauge fields as $A_{\mu}$ with $\mu=0,1,2$ , the electric field $E_i$ and the magnetic field $B$ are,
\begin{equation}
    E_i=\frac{\partial A_i}{\partial t}-\frac{\partial A_0}{\partial x^{i}} \,\text{ and }\, B=\frac{\partial A_2}{\partial x^1}-\frac{\partial A_1}{\partial x^2}
\end{equation}
and similarly for the other set of gauge fields $(\tilde{E}_i,\tilde{B})$. Using Bianchi identity $\epsilon^{\mu\nu\rho}\partial_{\mu}\partial_{\nu}A_{\rho}=0 $,
one can associate the conserved charges \eqref{concharge} and currents with those associated to these electric and magnetic fields. Upon identifications, one gets the following dictionary, 
\begin{eqnarray}
    && B=h \,,\qquad E_i=\epsilon_{ij}hu^j;
    \label{eq:4}\\
    && \tilde{B} = \zeta+f \,,\qquad \tilde{E}_i = \epsilon_{ij}(\zeta+f)u^j.
\end{eqnarray}
As can be seen, they are not fully independent and therefore redundant. Hence, all degrees of freedom of the shallow water system can be described by a single gauge field only. An action was introduced in  \cite{Tong:2022gpg} to get the correct EOM for the gauge theory
\begin{equation}
S=\int d^3x\left( \frac{E^2}{2B}-\frac{1}{2}gB^2+fA_0-\epsilon^{\mu\nu\rho}A_{\mu}\partial_{\nu}\beta\partial_{\rho}\alpha\right)
     \label{eq:5}
\end{equation}
The two scalar fields $\alpha$ and $\beta$ can be written upon considering the auxiliary gauge field as $\Tilde{A}_{\mu}$, using which  the last term in action \eqref{eq:5} reduces to the form of a Chern-Simons term, \footnote{This parameterisation (called a Clebsch parameterisation) is in general non-unique, hence the term is not a true Chern-Simons one in the nonlinear theory.}
\begin{equation}
\epsilon^{\mu\nu\rho}A_{\mu}\partial_{\nu}\beta\partial_{\rho}\alpha=\epsilon^{\mu\nu\rho}A_{\mu}\partial_{\nu}\Tilde{A}_{\rho}, \, \Tilde{A}_{\mu}=\partial_{\mu}\chi+\beta\partial_{\mu}\alpha.
\end{equation}
The action has gauge symmetries given by $A_{\mu}\rightarrow A_{\mu}+\partial_{\mu}\Lambda$ and $\Tilde{A}_{\mu}\rightarrow \Tilde{A}_{\mu}+\partial_{\mu}\Tilde{\Lambda}$ where $\Lambda$, $\Tilde{\Lambda}$ are gauge parameters.
EOM for $A_0$ is the Gauss' law which reads:
\begin{equation}
    \frac{\partial}{\partial x^i}\left(\frac{E_i}{B}\right)+f-\epsilon^{ij}\partial_i\beta\partial_j\alpha=0 \Rightarrow \text{ } \epsilon^{ij}\partial_i\beta\partial_j\alpha=\zeta+f.
\end{equation}
EOM for the spatial gauge field $A_i$ is 
\begin{equation}
     \frac{\partial}{\partial t} \left(\frac{E_i}{B}\right)+\frac{1}{2}\epsilon^{ij}\frac{\partial}{\partial x^j}\left(\frac{E^2}{B^2}\right)+g\epsilon_{ij}\frac{\partial B}{\partial x^j}
     -\epsilon_{ij}\tilde{E}_j=0.
\end{equation}
This equation when translated back to the fluid variables using \eqref{eq:4}, reduces to the second non-linear equation in \eqref{eq:1} and upon identifications using the Bianchi identity we get the first one in \eqref{eq:1}.

\begin{figure}[t]
	\centering
\includegraphics[width=9cm]{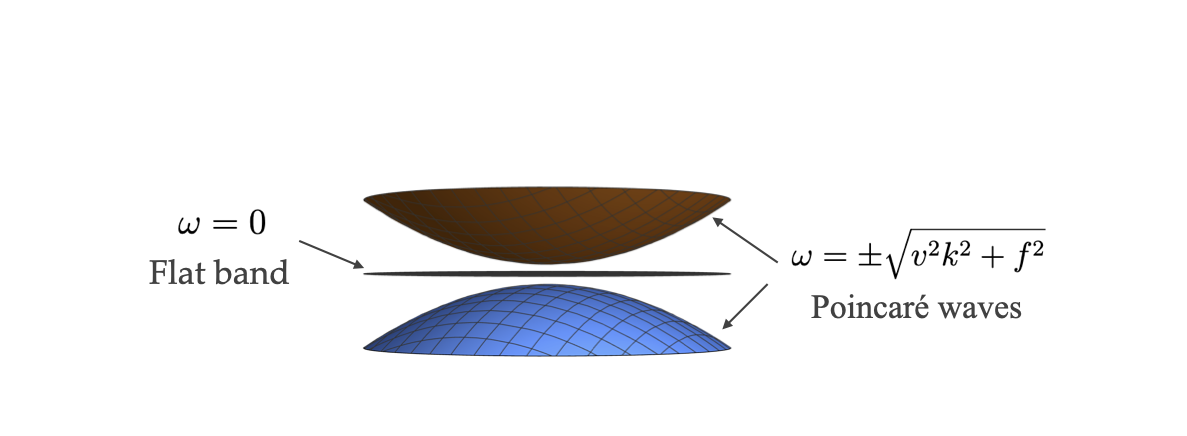}
\caption{Dispersion regimes of the shallow water solutions. }
\label{disp}
\end{figure}

\paragraph{Linearised theory.}
We start by linearising the theory of shallow water \eqref{eq:1}, by assuming that the variations of $h,u^i$ are small. This yields  $h(x^j,t)=H+\eta(x^j,t)$ and $u^i(x^j,t)=0+u^i(x^j,t)$, where $H$ is a constant height and $\eta \ll H$ \cite{Sheikh-Jabbari:2023eba}. Retaining terms up to linear order in the variations, the linearised shallow water equations read: 
\begin{eqnarray} \label{6}
    \frac{\partial \eta}{\partial t}+H\nabla.\Vec{u}=0, \quad \frac{\partial u^i}{\partial t}=f\epsilon^{ij}u^j-g\frac{\partial \eta}{\partial x^i}. \label{eq:6b}
\end{eqnarray}
To understand this theory better, we can look for solutions of the actual fluid equation (\ref{6}). We start with the following ansatz
\begin{equation}
    u_i=\hat{u}_ie^{i(\omega t-k.x)} \quad\text{and   } \quad \eta=\hat{\eta}e^{i(\omega t-k.x)}.
\end{equation}
Substituting them into the linear shallow water equations and solving the eigenvalue problem, we get solutions having two distinct dispersion relations (see Fig. \eqref{disp}):
\begin{equation}\label{omega}
    \omega=\pm \sqrt{v^2k^2+f^2} \quad\text{and}\quad \omega=0.
\end{equation}
The first one of these is known as Poincar\'{e} waves, and they have a relativistic-like dispersion with an effective mass $f$ and characteristic speed $v^2=gH$. For long wavelengths ($k\rightarrow 0$), these waves have a definite frequency fixed by the Coriolis parameter. For short wavelengths ($v|\Vec{k}|\gg f$), these waves travel with speed $v$. 
The second class of solutions with $\omega=0$, called flat-bands, indicate that there exists additional, time independent equilibrium solutions beyond the trivial solution $h=$ constant. These solutions have a non-trivial spatial profile with gravitational force balanced by a corresponding velocity profile $fu=-\partial h/\partial y$, which in turn generates a Coriolis force. This is known as \textit{geostrophic balance}.

Let us now in parallel consider the linearisation of the associated gauge theory with small fluctuations around the single set of gauge fields
\begin{equation}
    A_\mu=\hat{A}_{\mu}+\delta A_{\mu} \text{ with } \hat{A}_0=0, \partial_1 \hat{A}_2-\partial_2 \hat{A}_1=H.
\end{equation}
The fluid variables in terms of gauge fields become 
\begin{equation}
    B=\eta \quad\text{ and }\quad E_i=H\epsilon_{ij}u^{j}.
    \label{eq:14}
\end{equation}
Equation \eqref{eq:6b} in terms of fluid variables can then be transformed to:
\begin{equation}
    \Dot{E_i}=\epsilon_{ij}\left(fE_j-gH\partial_j B\right).
    \label{eq:10}
\end{equation}
Using the above, one can write the linearised effective gauge theory corresponding to both flat dispersion relation and the Poincaré-like dispersion relation.
These two cases give rise to different classes of actions, as would be elucidated next.

\paragraph{Effective action for flat-band.}
As discussed earlier, geostrophic flat bands in the fluid theory consist of non-trivial time-independent solutions of (\ref{eq:10}), given by $A_0=-\frac{v^2}{f}B$. Using this one can solve for the Gauss' law constraint, which takes the form $\partial_i E_i=\frac{v^2}{f}\nabla ^2B$. Following this an effective action for the flat band was proposed: 
\begin{equation}
    S_{geos}=\int dtd^2x \frac{1}{2H}\left(E_i-\frac{v^2}{f}\partial_iB\right)^2.
    \label{eq:13}
\end{equation}
The EOMs require that $E_i-(v^2/f)\partial_iB$ is constant, reproducing the geostrophic balance condition.
{To understand this, consider the equation \eqref{6} and let the fluid velocity be time-independent. Using the dictionary \eqref{eq:14} this leads to 
\begin{eqnarray}
	\partial_t u_i = 0 &&\Rightarrow \partial_t \left(\frac{1}{H}\epsilon_{ij}E^j\right) = 0 \Rightarrow \epsilon^{ij}\partial_t E_j = 0.
\end{eqnarray}
This can be solved when one considers either both $E^i$'s vary with time in the same way or $E^i$'s don't vary with time at all. The latter one i.e. $\partial_t E_i = 0$ is a peculiar kind of electrodynamic theory, that certainly does not have the usual symmetries of the relativistic Maxwell electric field.
The RHS of \eqref{eq:6b} in terms of gauge invariant observables translates to 
\begin{eqnarray}
	f\epsilon^{ij}u^j-g\frac{\partial \eta}{\partial x^i} 
	&& = \frac{f}{H}\left(E_i - \frac{v^2}{f} \partial_i B\right).
\end{eqnarray}
In the case above, the above quantity is actually zero which is nothing but the equation of motion for the flat band. }

\paragraph{Effective Action for Poincaré waves.} For the case of Poincaré waves, a gauge $A_0=0$ is chosen. The linearized action in this case can be written as
\begin{equation}\label{CSpoincare}
    S=\int dtd^2x \frac{1}{2H}\left(\Dot{A_i}^2-v^2B^2+f\epsilon^{ij}A_i\Dot{A_j}\right).
\end{equation}
Here 
the Gauss' law constraint takes the form $\partial_i E_i=fB$. One can reinstate $A_0$ in the action so that it reproduces the Gauss' law and the action which describes the dynamics of Poincaré waves is given by
\begin{equation}\label{CSfull}
        S_{Poinc}=\int dtd^2x \, \frac{1}{2H}\left(E^2-v^2B^2+f\epsilon^{\mu\nu\rho}A_\mu\partial_\nu A_\rho\right).
\end{equation}
This is the familiar relativistic Maxwell-Chern-Simons action, albeit defined with the characteristic velocity $v$. We assume an effective Poincaré invariant metric $g_{\mu\nu} = \text{diag}(-v^2,1,1)$, and the characteristic speed $v^2 = gH$. 
One can now find out the EOM associated to \eqref{CSpoincare}, which has an interesting form:
\begin{equation}\label{eqnPoinc}
	\partial_t\left(\dot{A_i} - f\epsilon^{ij}A_j\right) = -\epsilon^{ij}\partial_j\hat{B}, \quad \text{with} \, \, \hat{B} =vB,
\end{equation}
which looks like a 2+1 dimensional version of Faraday's law, if one identifies $\left(\dot{A_i} - f\epsilon^{ij}A_j\right)$ as a reimagined Electric field, but clearly that is not possible in the relativistic paradigm owing to symmetry considerations. 
We will now go ahead and try to connect both of these above cases to different avatars of theories with {Carrollian symmetry}, more concretely to Carrollian Electrodynamics. 

\paragraph{A quick look at Carroll symmetry.}

As introduced earlier, numerous counter-intuitive things occur in the Carroll world. The light cone closes completely, the space-time metric becomes degenerate, and new non-Lorentzian structures emerge \cite{Henneaux:1979vn}. Let us remind the reader that one can obtain the Carroll group by performing a contraction of the Poincare group, by taking the speed of light going to zero limit or equivalently
\begin{equation}
    x^i\rightarrow x^i \text{ , } t\rightarrow \epsilon t \text{ , } \epsilon \rightarrow 0.
\end{equation}
Under this limit, the Carroll generators are redefined by re-scaling the parent Poincare generators in the following way
\begin{equation}
    H \equiv \epsilon P_0 ^P , P_i \equiv P_i^P , C_i \equiv \epsilon J_{0i}^P , J_{ij} \equiv J_{ij}^P 
\end{equation}
where superscript $P$ refers to the Poincare generators, and the explicit form of the Carroll generators are:
\begin{equation}
    H=\partial_t , P_i=\partial_i , C_i=x_i\partial_t , J_{ij}=x_i\partial_j-x_j\partial_i
\end{equation}
where $H,P_i,C_i,J_{ij}$ are time and spatial translations, Carroll boosts and spatial rotations, respectively. The Lie algebra of the Carroll group is given by the following non-zero commutation relations:
\begin{subequations}
\begin{eqnarray}
    &&[J_{ij},J_{kl}]=4\delta_{[i[k}J_{l]j]},[J_{ij},P_k]=2\delta_{k[j}P_{i]}, \\
      && [J_{ij},C_k]=2\delta_{k[j}C_{i]},[C_i,P_j]=-\delta_{ij} H  
\end{eqnarray}
\end{subequations}
where $i,j,k,l=1,2,...,d$. Interestingly, Carroll boosts commute: $[C_i, C_j]=0$, making this a crucial difference from Poincaré algebra and indicating the non-Lorentzian nature of the algebra. 

\paragraph{Carrollian Electrodynamics.} 
We focus on a Carrollian version of Maxwell's theory $d=2+1$ dimension. Our method of construction closely follows the $c$-expansion of relativistic actions as elucidated in \cite{deBoer:2021jej}. The Maxwell field $A_\mu$ transforms under Lorentz boosts as 
\begin{equation}
    \delta A_\mu=ct{\beta}^k{\partial_k}A_\mu+\frac{1}{c}{\beta}^k{x}_k\partial_t A_\mu+\tilde{\delta}A_\mu
\end{equation}
where $\tilde{\delta}A_0={\beta}^k{A}_k$ and $\tilde{\delta}{A}_k={\beta}_kA_0$. We expand the Maxwell field in even powers of $c$ such that 
\begin{equation}
    A_\mu=c^\delta\left(A_\mu ^{(0)}+c^2 A_\mu ^{(1)}+...\right)
    \label{eq:11}
\end{equation}
for some $\delta$. We define ${\beta}_k=c{b}_k$ with ${b}_k$ the Carroll boost parameter along $k$-th direction. The field in the expansion in (\ref{eq:11}) then transforms as 
\begin{subequations}
\begin{eqnarray}
    &&\delta A_t ^{(n)}={\beta}^k{x}_k\partial_t A_t ^{(n)}+t{\beta}^k{\partial_k}A_t^{(n-1)}+{b}^k{A}_k^{(n-1)}\\
    &&\delta A_i ^{(n)}={\beta}^k{x}_k\partial_t A_i ^{(n)}+t{\beta}^k{\partial_k}A_i^{(n-1)}+b_iA_t^{(n)}.
\end{eqnarray}
Note that we have used $A_t=cA_0$. 
\end{subequations}
Equipped with the expansion of fields, we will now proceed to study the expansion of Maxwell Lagrangian $\mathcal{L}=-\frac{1}{4}F^{\mu\nu}F_{\mu\nu}$ where $F_{\mu\nu}=\partial_\mu A_\nu-\partial_\nu A_\mu$. Plugging in the expansion of the vector field, we can find that the Lagrangian expands according to $\mathcal{L}=c^{2\Delta-2}(\mathcal{L}_0+c^2\mathcal{L}_1+...)$ where the leading order Lagrangian is (see also \cite{Basu:2018dub}) 
\begin{equation}
    \mathcal{L}_0=\frac{1}{2}\left(F_{0i}^{(0)}\right)^2 =\frac{1}{2}\left(E_i^{(0)}\right)^2\qquad i=1,2\label{ced1}\\
\end{equation}
with $F_{0i}=E_i^{(0)}=\partial_tA_i^{(0)}-\partial_iA_t^{(0)}$. This leading order Lagrangian $\mathcal{L}_0$ is called the Electric Carroll Electrodynamics Lagrangian and this is naturally Carroll invariant. The next to leading order Lagrangian is 
\begin{equation}
    \mathcal{L}_1=F_{0i}^{(0)}F_{0i}^{(1)}-\frac{1}{4}\left(F_{ij}^{(0)}\right)^2 =E_i^{(0)}E_i^{(1)}-\frac{1}{2}\left(B^{(0)}\right)^2 \label{ced2}
\end{equation}
where $F_{ij}^{(0)}=B^{(0)}=\partial_i A_j^{(0)}-\partial_j A_i^{(0)}$ and $F_{0i}^{(1)}=E_i^{(1)}=\partial_tA_i^{(1)}-\partial_iA_t^{(1)}$.
It should be noted here that though $\mathcal{L}_0$ is Carroll invariant, the subleading $\mathcal{L}_1$ is not. Specifically Carroll boosts are not a symmetry of $\mathcal{L}_1$. In order to restore Carroll symmetry at the subleading order, we introduce a constraint that implements the leading order equations of motion. This modified Lagrangian is called the Magnetic Lagrangian, which now is Carroll invariant (see also the discussion in \cite{Henneaux:2021yzg}). More details of this can be found in the supplementary material. 

\paragraph{From Carroll to water waves: Flat bands.} Now for our magic tricks. We have already seen earlier that the flat band effective action gives rise to a time independent electric field EOM. From \eqref{ced1}, the leading order action is
\begin{equation}\label{la1}
    S^{(0)}=\int d^3x\,\frac{1}{2}\left(\partial_t A_i^{(0)}-\partial_i A_t^{(0)}\right)^2.
\end{equation}
Recall the action for the flat-band of shallow water waves in \eqref{eq:13}.
Rather remarkably we see that we can now map the Carrollian ED action to the flat band action with the help of the following identifications 
\begin{equation}\label{gidenti}
	A_t^{(0)} = \frac{1}{\sqrt{H}}\left(A_0 + \frac{v^2}{f}B\right),~~~
 A_i^{(0)} = \frac{1}{\sqrt{H}}A_i.
\end{equation}
where the RHS indicate the fields in the flat-band action (\ref{eq:13}). This means 
\begin{equation}
E_i^{(0)} = \frac{1}{\sqrt{H}} \left({E_i}-\frac{v^2}{f}\partial_i B\right).
\end{equation}
So we have an exact mapping between two apparently unrelated systems. 

 Note that the Electric Carroll action \eqref{la1} is invariant under the gauge transformations
\begin{equation}
	A_t^{(0)} \to A_t^{(0)} + \partial_t \lambda_1\,,\quad A_i^{(0)} \to A_i^{(0)} + \partial_i \lambda_2
\end{equation}
provided $\lambda_1 - \lambda_2 = \text{spacetime constant}$. Thus one could show \eqref{gidenti} itself is a gauge invariant identification.
The equations of motion of \eqref{la1} read
\begin{equation}
   \partial_i(\partial_t A_i^{(0)}-\partial_i A_t^{(0)})=0 \text{ , } \partial_t(\partial_t A_i^{(0)}-\partial_i A_t^{(0)})=0
\end{equation}
Using the identification, we can write these as
\begin{subequations}
\begin{eqnarray}
    &&\frac{1}{\sqrt{H}}\partial_i \left({E_i}-\frac{v^2}{f}\partial_i B\right)_{\text{flat band}}=0 \\
    &&\frac{1}{\sqrt{H}}\partial_t \left({E_i}-\frac{v^2}{f}\partial_i B\right)_{\text{flat band}}=0.
\end{eqnarray}
\end{subequations}
Translating back to the fluid variables using the dictionary in (\ref{eq:14}), we have for the first equation,
\begin{equation}
    H\epsilon^{ij}\partial_i u_j=\frac{v^2}{f}\nabla^2\eta,
\end{equation}
which, indeed, can be found using the shallow water equations in (\ref{eq:6b}) for flat-band case (time-independent solutions). 

Using the Bianchi identity in $d=3$ and the identification \eqref{gidenti} we have
$\epsilon^{ij}\partial_i E_j =0$, or equivalently  $\nabla.\Vec{u}=0$, 
which is indeed true for an incompressible fluid. Thus we see that our mapping is justified.
Looking at the energy-momentum tensor, we find that 
\begin{equation}
        T^i_t=\frac{1}{2H}\left(E_i-\frac{v^2}{f}\partial_iB\right)\partial_t A_t.
\end{equation}
Using the flat band condition $E_i=\frac{v^2}{f}\partial_iB$, we see that $T^i_t=0$, which is a tell-tale signature of Carroll boost invariance.

\paragraph{From Carroll to water waves: Poincaré Waves.}

Now we consider Poincaré waves. 
Remember, in the $A_0=0$ gauge, we can write the Chern-Simons derived action \eqref{CSpoincare}.
Also recall the subleading order action for Carroll electrodynamics from \eqref{ced2} in $d=2+1$,
\begin{equation}
    \mathcal{L}_1=E_{i}^{(1)}\left(\partial_tA_i^{(0)}-\partial_iA_t^{(0)}\right)-\frac{1}{2}\left(\epsilon^{ij}\partial_iA_j^{(0)}\right)^2 .
\end{equation}
 Here also we pick a similar gauge $A_t^{(0)}=0$ and the action becomes
\begin{equation}\label{PWaction}
     S^{(1)}=\int dtd^2x \left(E_{i}^{(1)} \partial_t A_i^{(0)} -\frac{1}{2}(B^{(0)})^2\right).
\end{equation}
We will now see that Poincaré waves can be thought of the leading fluctuation over the geostrophic flat-band solution. To do this, we match the subleading order electrodynamics action obtained above to the Poincaré waves action
\footnote{It could be shown that both (\ref{PWaction}) and (\ref{CSpoincare}) have no residual gauge freedom left. See \cite{Banerjee:2020qjj} for comments on gauge symmetries in Magnetic Carroll electrodynamics.}.
Using the identifications (see \eqref{eqnPoinc})
\begin{equation}
  A_i^{(0)}= \frac{A_i}{\sqrt{2H}}  \,\text{ , }\, B^{(0)}= \pm \hat{B},
  \end{equation}
a simple comparison of \eqref{PWaction} and \eqref{CSpoincare} leads to 
\begin{equation}\label{psi}
    E_{i}^{(1)} \equiv \partial_t A_i^{(0)}-f\epsilon^{ij}A_j^{(0)}.
\end{equation}
The equation of motion for $A_i^{(0)}$ derived from $S^{(1)}$ reads exactly the same as in \eqref{eqnPoinc} :
\begin{equation}
\partial_t E_i^{(1)} = - \epsilon^{ij} \partial_jB^{(0)}.
\end{equation}
Evidently, as we plug the identification back in, we get the following action:
\begin{equation}
     S^{(1)}=\int dtd^2x \left((\partial_t A_i^{(0)})^2-(B^{(0)})^2+f\epsilon^{ij}A_j^{(0)}\partial_t A_i^{(0)}\right)
\end{equation}
which is in the same form as in \eqref{CSpoincare}.

At this juncture, several important questions arise, the first and foremost among them would be about symmetries. The action proposed in \cite{Tong:2022gpg} is just a relativistic Maxwell Chern-Simons action which clearly has relativistic symmetries. So, how and why do Carroll symmetries arise in this context at all? 

To understand this, we need to remind the reader of the discussion below \eqref{ced2}. The subleading Carroll Magnetic theory is also Poincare invariant at first glance, but one needs to impose the validity of leading order equations of motion as a constraint on the system to see that Carroll symmetries emerge. In the case of shallow water, the argument stays the same. The Poincare waves are fluctuations on top of the flat band. Drawing analogies with e.g. gravitational waves on a certain spacetime or the background field method in quantum field theory, the background for the Poincare wave is that of the flat band. The equations of motion of the flat band have to hold in this case. This translates to $E_i^{(0)}=0$. It can be explicitly checked that once this constraint is imposed, the action transforms as \footnote{See the supplement for some details.} 
\begin{equation}
    \delta \mathcal{L}_1=b^kx_k\partial_t \mathcal{L}_1.
\end{equation}
So we have invariance under Carroll symmetry. It can also be checked that the $T^i_{\, t}$ component of the stress tensor for this case is zero once the leading order equations of motion are imposed. This, as we emphasised before, is a signature of Carroll boost invariance and more generally indicates the emergence of Carrollian symmetry in the system.

We have found that both the flat band and the Poincaré wave have an underlying Carroll symmetry. It is important to emphasise here that not only are the actions invariant under the finite Carroll algebra introduced earlier, but they are invariant under an infinite dimensional version which includes the so called supertranslations: $M(x^i) = f(x^i) \partial_t$ where $f(x^i)$ is an arbitrary function of the spatial coordinates $x^i$. 

\paragraph{Discussions.}
In this letter, we put forward an intriguing connection between two seemingly disparate theories, that of linearised shallow water waves and Carroll Electrodynamics. By identifying the gauge theoretic description of geostrophic flat bands and Poincaré waves given in \cite{Tong:2022gpg} to explicitly electric and magnetic Carroll Electrodynamics, we establish a dictionary between Carroll fields and shallow fluid variables for the two physical regimes. 

We should emphasise how counter-intuitive this connection is at the outset. Shallow water waves constitute a very non-relativistic situation while Carrollian symmetries generally occur when things move at or near the speed of light. A priori having an ultra-relativistic symmetry appear in a non-relativistic context seems nonsensical. But we have explicitly showed this by mapping actions. Now let us offer an explanation as to why this seemingly contradictory situation is actually natural. 

\begin{figure}[t]
	\centering
\includegraphics[width=9cm]{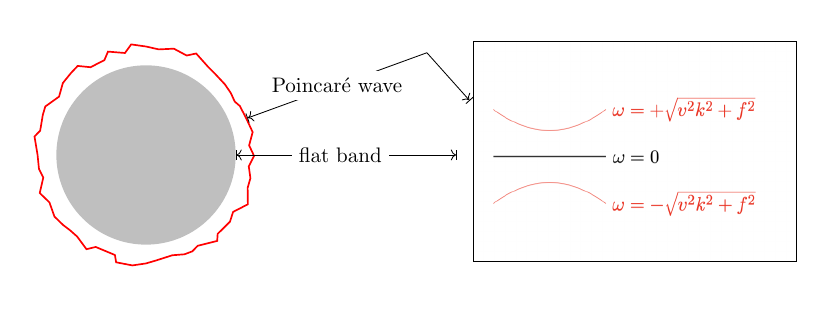}
\caption{A heuristic depiction of the acoustic black hole. In the left flat-band (grey, on the horizon) and Poincaré wave (red, fluctuations around horizon) solution identified with the dispersion relation in the right.}
\label{inset}
\end{figure}

As mentioned in the introduction, Carrollian symmetries arise on generic null surfaces and specifically on the event horizon of black holes \cite{Donnay:2019jiz}. There exists an interesting formulation of shallow water waves in terms of analogue gravity models utilising the Painlevé-Gullstrand effective metric (see e.g. \cite{Faccio:2013kpa}). This metric has a horizon depicted by the vanishing of its $g_{tt}$ component which corresponds precisely to the vanishing of $\omega$ (see \eqref{omega}). The flat band solution thus sits on the horizon of the acoustic analogue metric, i.e. variables on the RHS of \eqref{gidenti} are defined \textit{on} this acoustic horizon. It is thus very natural that the flat band solution has an underlying Carrollian symmetry. It has been recently also seen that the near horizon region of black holes \cite{Bagchi:2023cfp} and Milne horizons \cite{Bagchi:2023ysc} also come equipped with Carroll structures. It is also expected that the Poincaré waves, which are departures from flat band, sit in the near-horizon of the acoustic metric and also inherit a Carroll structure. See Fig. \eqref{inset} for a depiction. 

Furthermore, in a $c$ expansion, the Electric limit is the leading one and the Magnetic is the subleading one, as we have seen earlier. In case of a black hole, the distance from the horizon acts as an effective speed of light. So theories that sit directly on the horizon would correspond to the leading order theory and the ones in the near-horizon region, slightly away from the horizon, would constitute the subleading theory. From this point of view, it makes sense that the Electric version of Carroll electrodynamics corresponds to the flat band which lies directly on the acoustic horizon and the Poincaré waves are given by the Magnetic version. It would be instructive to explicitly show the emergence of Carroll electrodynamics from the relativistic Chern-Simons theory \eqref{CSfull} incorporating the ideas of the acoustic black hole that we have alluded to above. We hope to report on this in the near future. 

Going forward, let us comment on another potentially very interesting direction. In the recent studies of the infrared region of gauge theories and gravity, a rather wonderful connection between asymptotic symmetries and memory effects have been discovered \cite{Strominger:2014pwa}. It has been understood that supertranslations in asymptotically flat spacetimes is responsible for the gravitational memory effect. The vacuum in general relativity is highly degenerate and these vacua are related to each other by the addition of soft gravitons. Passage of radiation through a region induces a transition between these vacua related by supertranslations \footnote{See \cite{Strominger:2017zoo} for a wonderful review.}. The observable effect, called the memory effect, is the relative spatial displacement of nearby detectors when radiation passes through. Thus the memory effect is directly related to supertranslations. Memory effects have been found in shallow water waves in \cite{Sheikh-Jabbari:2023eba} and we have now discovered supertranslation symmetry directly associated to these waves. It is very tempting to construct the analogue of the above connection between these two in the context of shallow water waves. This is work in progress.

Finally, since a gauge theory description of such shallow water waves are important in many physical phenomena, including solar tacholine and Magnetohydrodynamics \cite{Lier:2024svb}, construction of solitonic waves \cite{Zabusky_Galvin_1971} etc., one could view the current work merely as a prelude to fully utilising the power of Carroll symmetries to understand these physical situations better. We will report more on these problems elsewhere.


\bigskip

\noindent \textbf{Acknowledgements.} We thank Rudranil Basu, Kedar Kolekar, Hisayoshi Muraki, Ashish Shukla, Shahin Sheikh-Jabbari  and particularly David Tong for useful discussions and comments on the manuscript. 

AB is partially supported by a Swarnajayanti Fellowship from the Science and Engineering Research Board
(SERB) under grant SB/SJF/2019-20/08 and also by an ANRF grant CRG/2022/006165. ABan is supported in part by an OPERA grant and a seed grant NFSG/PIL/2023/P3816 from BITS-Pilani. He also acknowledges financial support from the Asia Pacific Center for Theoretical Physics (APCTP) via an Associate Fellowship. SM is supported by grant number 09/092(1039)/2019-EMR-I from Council of Scientific and Industrial Research (CSIR). SS is supported by an IIT Kanpur Institute Assistantship.


\bibliographystyle{utphys2}
\bibliography{ccft}

\bigskip

\section*{Supplementary material}
\subsection*{Carroll boost invariance for Poincaré waves}
As mentioned in the main text, action for the Poincaré waves can be written by identifying $\left(\dot{A_i} - f\epsilon^{ij}A_j\right)$ as a changed electric field $E_i'$: 
\begin{equation}
	\mathcal{L}_{poinc} = \frac{1}{2H}\left(E_i'\partial_tA_i - v^2B^2\right)
\end{equation}
Boost transformation rules for the relativistic fields in $2+1$ dimensions are given by
\begin{eqnarray}
	&&\delta A_i=ct{\beta}^k{\partial_k}A_i+\frac{1}{c}{\beta}^k{x}_k\partial_t A_i+ \beta_iA_0 \nonumber \\
	&&\delta E_i' = ct{\beta}^k{\partial_k}E_i'+\frac{1}{c}{\beta}^k{x}_k\partial_t E_i'+ (\beta_i - f\epsilon^{ij}\beta_j)A_0 + c{\beta}^k{\partial_k}A_i \nonumber\\
	&&\delta B = ct{\beta}^k{\partial_k}B+\frac{1}{c}{\beta}^k{x}_k\partial_t B + \epsilon^{ij}\beta_j\partial_i A_0 + \frac{1}{c}\beta^i\epsilon^{ij}\partial_tA_j
\end{eqnarray}
under Carroll limit $(\beta = cb, A_t = cA_0, A_i =A_i)$ and $c\to 0$, equations above become:
\begin{eqnarray}
	&&\delta A_i=b^kx_k\partial_t A_i+ b_iA_t \nonumber \\
	&&\delta E_i' = b^kx_k\partial_t E_i'+ (b_i - f\epsilon^{ij}b_j)A_t  \nonumber \\
	&&\delta B = b^kx_k\partial_t B + \epsilon^{ij}b_j\partial_i A_t +b^i\epsilon^{ij}\partial_tA_j
\end{eqnarray}
Choosing a gauge $A_t = 0$, and combining all, we get:
 \begin{eqnarray}
	\delta \mathcal{L}_{poinc} = b^kx_k\partial_t \mathcal{L}_{poinc} - 2v^2Bb^i\epsilon^{ij}\partial_tA_j
\end{eqnarray}
The action will be Carroll boost invariant provided $\dot{A_i}=0$. This fact can be mapped one-to-one to the magnetic Carroll electrodynamics which is not invariant under Carroll boots but can be made so by explicitly putting $E_i^{(0)} = 0$. 

The emergence of Carrollian symmetry can be similarly seen by constructing the the stress tensor and focussing on the $T^{i}_{\,\,t}$ component:
\begin{equation}
	T^{i}_{\,\,t} =  \frac{\partial \mathcal{L}_{poinc}}{\partial_iA_t}\partial_tA_t
 + \frac{\partial \mathcal{L}_{poinc}}{\partial_iA_j}\partial_tA_j
\end{equation}
 the first term identically vanishes as there is no $A_t$ field. The last term vanishes upon imposition of the constraint $\dot{A}_i=0$.  This again solidifies emergence of Carroll boost in this theory.

\end{document}